\documentclass[11pt]{article}
\usepackage{amssymb}
\usepackage{amsmath}
\usepackage{graphicx}
\usepackage{setspace}
\usepackage{fancyhdr}
\usepackage{ifpdf}
\usepackage{graphicx}
\usepackage{comment}

\def\half{{\textstyle{1\over2}}}

\def\G{\Gamma}

\newcommand{\bea}{\begin{eqnarray}}
\newcommand{\eea}{\end{eqnarray}}
\newcommand{\be}{\begin{equation}}
\newcommand{\ee}{\end{equation}}

\newcommand{\bi}{\begin{itemize}}
\newcommand{\ei}{\end{itemize}}

\newcommand{\qed}{\nobreak \ifvmode \relax \else
      \ifdim\lastskip<1.5em \hskip-\lastskip
      \hskip1.5em plus0em minus0.5em \fi \nobreak
      \vrule height0.75em width0.5em depth0.25em\fi}

\def\ben{\begin{equation}}
\def\een{\end{equation}}
\def\half{{\textstyle{1\over2}}}

 \let\t=\tau

\let\w=\omega \let\G=\Gamma \let\D=\Delta

\let\pa=\partial
\def\be{\begin{equation}}
\def\ee{\end{equation}}
\def\beq{\begin{equation}}
\def\eeq{\end{equation}}
\def\ba{\begin{array}}
\def\ea{\end{array}}

\def\dalemb#1#2{{\vbox{\hrule height .#2pt
       \hbox{\vrule width.#2pt height#1pt \kern#1pt
               \vrule width.#2pt}
       \hrule height.#2pt}}}

\def\R{{{\Bbb R}}}

\def\Z{{{\Bbb Z}}}

\def\N{{{\Bbb N}}}

\allowdisplaybreaks  

\begin{document}

\thispagestyle{empty}

\begin{flushright} NSF-KITP-11-202 \end{flushright}

\vspace*{1cm}
\begin{center}
{ \Large {\textsc{Static Patch Solipsism: \\ Conformal Symmetry of the de Sitter Worldline}}}
\vspace*{1.5cm}
 
Dionysios Anninos$^{\flat}$, Sean A. Hartnoll$^{\flat\sharp}$ and Diego M. Hofman$^\sharp$ 
\vspace*{1cm}

{\it $^\flat$ Department of Physics, Stanford University, \\ Stanford, CA 94305-4060, USA}
\vspace*{0.5cm}

{\it $^\sharp$ Center for the Fundamental Laws of Nature, \\ Harvard University, Cambridge, MA 02138, USA}
\end{center}
\vspace*{1.2cm}

\begin{abstract}

We show that the propagators of gravitons and scalar fields seen by a static patch observer in de Sitter spacetime are controlled by hidden $SL(2,\R)$ symmetries, at all frequencies.  The retarded Green's function is determined by an $SL(2,\R) \times SL(2,\R)$ action generated by conformal Killing vectors of de Sitter spacetime times a line. This observation uses the fact that the static patch of dS$_{d+1}\times \R$ is conformal to the hyperbolic patch of AdS$_3 \times S^{d-1}$. The poles of the propagators, the quasinormal frequencies, are generated by associated $SL(2,\R)$ actions. The quasinormal mode generating algebras capture the conformal weights more usually read off from the fields at future and past infinity. For conformally coupled scalar fields, and for gravitons in four dimensions, this $SL(2,\R)$ algebra has an enhanced supersymmetric structure and
is generated by particular conformal Killing vectors of de Sitter spacetime. We show how the worldline de Sitter propagators can be reproduced from a `level matched' left and right moving conformal quantum mechanics with an appropriate spectrum of primary operators. Our observations are consistent with the notion that the static patch of de Sitter spacetime is dually described by a (level matched) large $N$ worldline conformal quantum mechanics.

\end{abstract}
\newpage
\setcounter{page}{1}
\onehalfspacing

\section{Towards a solipsistic holography}

There are at least two types of observers in a de Sitter universe. The static patch observer is surrounded by a cosmological horizon and is immersed in a thermal bath at the de Sitter temperature \cite{Gibbons:1977mu}. The `metaobserver', in contrast, observes all of the future de Sitter boundary $\mathcal{I}^+$, e.g. via the wavefunction of the universe \cite{Hartle:1983ai}. Cosmologically speaking, we are approximately both a static patch observer, given that we are entering a new de Sitter phase, and also metaobservers of the past inflationary de Sitter phase of our universe. Thus it is important to understand both types of observers. The $\mathcal{I}^+$ metaobserver has been discussed extensively in holographic approaches to de Sitter space, including the dS/CFT correspondence \cite{Witten:2001kn, Strominger:2001pn, Maldacena:2002vr, Anninos:2010zf, Anninos:2011ui, arXiv:0907.5542}. In static patch holography the cosmological horizon, finite temperature, and reduced symmetry of the static observer are all expected to play an important role \cite{Goheer:2002vf, Banks:2003cg, Parikh:2004wh, Banks:2006rx, Susskind:2011ap, Castro:2011xb}. Other approaches to the static patch include de Sitter foliations of the patch \cite{Alishahiha:2004md} and their subsequent use to parametrically estimate the de Sitter entropy by realizing such foliations in flux compactifications \cite{Silverstein:2003jp, Dong:2010pm}.

This letter will study a simple aspect of static patch holography, the retarded Green's function of gravitational and scalar field fluctuations seen by a static patch observer. We will take several cues from the established AdS/CFT correspondence \cite{Maldacena:1997re,Witten:1998qj,Gubser:1998bc}. It has been argued that a dual description at the dS future boundary is the appropriate analogy to the AdS boundary description. However, the fact that the future boundary is behind the de Sitter horizon may suggest that a description at future infinity must necessarily grapple with the entire landscape of string theory, or at least the subset thereof which makes up the de Sitter entropy, perhaps involving ideas suggested in \cite{Freivogel:2006xu, Susskind:2007pv, Freivogel:2009rf}. It is useful to make an analogy with an eternal black hole in AdS space. Outside of the horizon, which includes the spatial boundary, the complicated physics of the black hole microstates are subsumed into the relatively simple description of a finite temperature quantum field theory. An analogous simple description in dS space can be expected for the static patch observer, while the metaobserver is more analogous to a putative holographic description of the black hole at the spacelike singularity (the presence or absence of a curvature singularity is not the issue here).

The static patch observer is not without its share of complications. In particular, the observer is in the midst of a gravitating, in general quantum and potentially unstable \cite{Ginsparg:1982rs} space-time and therefore does not define a superselection sector of the theory in the same way as an observer at the spatial boundary of AdS. Furthermore, the static patch observer breaks the $SO(1,d+1)$ symmetry of dS$_{d+1}$ down to $SO(d) \times \R$. The larger symmetry group of the full spacetime relates different static patch observers. In this paper we will take the `solipsistic' (i.e. gauge fixed) point of view that establishing the existence of other static observers is not necessary for a consistent description of a static patch. Developing an interpretational framework will not be necessary for the purposes of this paper; the general idea is that a possible dual large $N$ quantum mechanics would reproduce the measurements made by a static patch observer in de Sitter spacetime. It is of interest, however, to relate the (perturbative) data that we will construct in this letter, pertaining to a single static patch observer, to the perspective of the metaobserver at $\mathcal{I}^+$. We will partially do so, using some of the ideas developed in \cite{Bousso:2001mw,Anninos:2010gh,Anninos:2011jp}. Ultimately this may help to address the serious challenges faced in holographically reconstructing the Lorentizan static patch observer, even granted the knowledge of a dual Euclidean theory at $\mathcal{I}^+$. It may also connect the perspective of the solipsistic observer with ideas associated with complementarity in de Sitter space \cite{Bousso:2002ju}.

The main observation of this letter will be that the retarded Green's functions of gravitons and scalar fields for a static patch observer are those of `level matched' primary operators in a left and right moving conformal quantum mechanics. The static patch de Sitter time corresponds to evolution under the $H-K$ generator of the conformal quantum mechanics. The structure of the Green's functions is controlled by hidden $SL(2,\R) \times SL(2,\R)$ actions. The generators of these algebras originate from an AdS$_3$ spacetime that is conformally related to the original de Sitter space times a line, and may hint at the presence of an underlying CFT. Descending from this $SL(2,\R) \times SL(2,\R)$ algebra, we obtain a further $SL(2,\R)$ which generates the quasinormal mode spectrum of de Sitter spacetime. From the Casimir of this latter conformal algebra, we
can read off the weights of the de Sitter fields that are usually associated with the behavior of the fields at future and past infinity.
The `quasinormal mode algebra' has an enhanced supersymmetric-like structure for conformally coupled scalars and for gravitons in precisely four dimensions. In these cases the spectrum generating $SL(2,\R)$ is linearly generated by conformal Killing vectors of de Sitter spacetime.

\section{Propagators on the static patch worldline}
\label{sec:first}

The static patch coordinate system for $(d+1)$-dimensional de Sitter spacetime is
\be \label{eq:dsmetric}
{ds^2} = -f dt^2 + \frac{dr^2}{f} + r^2 d\Omega_{d-1}^2~,
\ee
where
\be
f = 1-(r/\ell)^2 \,,
\ee
with $r \in [0,\ell]$ and $d\Omega_{d-1}^2$ the round metric on $S^{d-1}$.
The static patch observer is at $r=0$ and the cosmological horizon, at $r=\ell$.
We start with the case of a scalar field. The wave equation for a scalar field of mass $m$ is
\be 
\frac{1}{\sqrt{-g}} \partial_\mu g^{\mu\nu} \sqrt{-g} \partial_\nu \Phi = m^2 \Phi~.
\ee
Separating variables using spherical harmonics $Y_{l}(\Omega)$,
\be
\Phi(t,r,\Omega) =  \varphi (r) e^{- i \omega t} Y_{l}(\Omega) \,,
\ee
the wave equation becomes
\be\label{eq:wave}
-\frac{1}{r^{d-1}} \frac{d}{dr} \left(f \, r^{d-1} \frac{d \varphi}{dr} \right) + m^2 \varphi + \frac{l (l + d-2)}{r^2} \varphi = \frac{\w^2}{f} \varphi \,.
\ee
We will concentrate on light scalar fields in the following and thereby write
\be\label{eq:mm22}
\ell^2 m^2 = \frac{d^2}{4} - x^2 \,.
\ee
Mostly we will have in mind $\frac{d}{2} > x > 0$. Taking $x$ larger corresponds to a negative mass squared scalar, while $x$ must be taken imaginary to describe heavier fields. We will note that many of our results hold also for the heavier fields with $x$ imaginary and in some contexts imaginary $x$ appears to be more natural.

The general solution to the wave equation has the form
\be\label{eq:gensol}
\varphi = B \varphi_\text{n.} + A \varphi_\text{n.n.} \,,
\ee
with the two solutions distinguished by their behaviour at the origin $r \to 0$:
\bea\label{eq:hyper}
\varphi_\text{n.} & = & \left(1- \frac{r^2}{\ell^2}\right)^{-i\omega/2} \left(\frac{r}{\ell}\right)^l  {}_2F_1\left(a + h_-, a + h_+ ; \frac{d}{2} + l; \frac{r^2}{\ell^2}\right)~, \\
\varphi_\text{n.n.} & = & \left(1- \frac{r^2}{\ell^2}\right)^{-i\omega/2}\left(\frac{r}{\ell}\right)^{2-d-l} {}_2F_1\left(b + h_-, b + h_+ ; \frac{4-d}{2} - l; \frac{r^2}{\ell^2}\right)~. \nonumber
\eea
Here $a = (l-i\ell \omega)/2$, $b=(2-d-l-i\ell \omega)/2$ and the weights
\be\label{eq:weights}
h_\pm = \frac{d}{4} \pm \frac{x}{2} \, .
\ee
Based on the falloffs near the origin, we can will call $\varphi_\text{n.}$ normalizable and $\varphi_\text{n.n.}$ non-normalizable. The two solutions are related by $l \leftrightarrow 2 - d - l$. A particularly interesting case will be conformally coupled scalars with $x = \half$ (in all dimensions).

Given the equations, we must impose boundary conditions defining the physical excitations.
Following our discussion in the introduction, emphasizing an analogy with the well understood AdS/CFT dualities, we would like to think of our dual degrees of freedom as living at the origin of the static patch, where the celestial sphere degenerates and where the static patch observer is located. The simplest objects to compute are the worldline retarded Green's functions as seen by the static patch observer. Inspired by the prescription for computing retarded Green's functions introduced by \cite{Son:2002sd} for the AdS/CFT case at finite temperature, we impose that the flux associated to the wave equation is ingoing on the future cosmological horizon of the static patch (i.e. going into the future horizon). This can equivalently be phrased as the absence of flux emanating from the past horizon. From the perspective of global de Sitter space there would be no a priori reason to impose this condition. However, solipsistically speaking, this choice allows us to study the response to an observer sending out isolated pulses into a de Sitter universe. Our retarded Green's functions in the static patch are related to de Sitter invariant Green's functions \cite{134659}, restricted to the worldline of an observer, by multiplication by a step function $\theta(t)$.

Expanding the hypergeometric functions in the solutions (\ref{eq:hyper}) near the horizon, as $r \to \ell$, one finds the two behaviors: $\varphi \sim \left(1-r^2/\ell^2 \right)^{\pm i\ell \omega/2}$. The ingoing boundary condition is then the requirement that, near the horizon, $\varphi$ has no terms proportional to $\left(1-r^2/\ell^2 \right)^{i\ell \omega/2}$. This condition linearly relates $A$ and $B$ in the general solution (\ref{eq:gensol}).

Again by analogy with the AdS/CFT case, we may expect that the non-normalizable mode can be thought of as the source and the normalizable mode as the response, and thereby anticipate that the retarded worldline Green's function will be proportional to the ratio $B/A$. Relating $A$ and $B$ through the ingoing boundary condition at the horizon, as described above, we find
\be
G^R_l(\w) \, \propto \, \frac{B}{A} \, \propto \,
\left(\frac{1}{\ell}\right)^{2l+d-2}\frac{\Gamma\left(2 - \half d - l \right)}{\Gamma\left(\half d + l \right)} P_+(\w) P_-(\w) \,, \label{eq:aa}
\ee
where
\be
P_\pm(\w) = \frac{\Gamma\left( \frac{1}{2}(l-i\ell \omega)+h_\pm\right)}{\Gamma\left( \frac{1}{2}(2-d-l-i\ell \omega)+h_\pm\right)} \,.
\ee
This formula must be interpreted with care in the case of odd spacetime dimension $d+1$. There, the gamma function in the numerator of (\ref{eq:aa}) diverges while the residues of the poles vanish. The correct Green's function in this case is obtained by analytically continuing to general noninteger $l$, taking the limit $l \to \N$, and neglecting a divergent term that is analytic in frequencies. The result is, dropping analytic contact terms,
\be
G^R_l(\w) \, \propto P_+(\w) P_-(\w) \sum_\pm \Big(\psi\left( \half (l-i\ell \omega)+h_\pm\right) + 
\psi\left( \half (2-d-l-i\ell \omega)+h_\pm\right)  \Big) \,.
\ee
This result should also follow from the wavefunctions constructed in \cite{LopezOrtega:2006my} for odd dimensions. The $P_\pm(\w)$ multiplying the digamma functions give a polynomial in $\w$.

Anticipating our discussion below, we can note that the Green's function (\ref{eq:aa}) simplifies for a conformally coupled scalar
\be
G^R_l(\w) \propto \left(\frac{1}{\ell}\right)^{2l+d-2} \frac{\Gamma\left(2 - \half d - l \right)}{\Gamma\left(\half d + l \right)} \frac{\Gamma\left(\half (d-1) + l - i \ell \w \right)}{\Gamma\left(\half (3-d) -l - i \ell \w \right)} \,, \qquad (x=\half) \,. \label{eq:Ggrav}
\ee
This formula needs to be interpreted with care for both even and odd dimensions. Odd dimensions proceed as above, but the even dimensional case is subtle here because the poles of the correct Green's function are precisely at the Matsubara frequencies (for temperature $T_\text{dS} = 1/(2\pi \ell)$).
The most concise way to make sense of (\ref{eq:Ggrav}) is to analytically continue to noninteger $l$ and compute the Fourier transform to obtain
\be\label{eq:time}
G^R_l(t) \, \propto \,  \theta(t) \left( \frac{\ell^{-1}}{\sinh \half t/\ell} \right)^{2 l + d-1} \,, \qquad (x= \half) \,.
\ee
In this expression we may safely take $l$ integer in any dimension and then Fourier transform back to frequency space if we so desire, to obtain an expression in terms of digamma functions. As we will recall below, (\ref{eq:time}) is precisely the correlator of an operator with weight
\be\label{eq:deltaA}
\Delta =  l + \half (d-1) \,,
\ee
in an $SL(2,\R)$ invariant quantum mechanics, evolved with respect to a specific time generator. The more general propagator (\ref{eq:aa}) Fourier transforms to the convolution of two propagators of the form (\ref{eq:time}). We will explain below how this structure arises from level matching a left and right moving $SL(2,\R)$.

We now prove that $B/A$ is indeed the worldline Green's function. This is easily done using the standard formulae for Green's functions as in e.g. \cite{Ching:1994bd}. Perform a coordinate transformation and rescale the wavefunction
\be
r = \ell \tanh y \,, \qquad \varphi = r^{(1-d)/2} \chi \,.
\ee
Then the wave equation (\ref{eq:wave}) is in Schr\"odinger form with a P\"oschl-Teller-like potential
\be\label{eq:schrodinger} 
- \frac{d^2 \chi}{dy^2} + \left(\frac{\Delta (\Delta - 1)}{\sinh^2 y} - \frac{(x + \half)(x - \half)}{\cosh^2 y} \right) \chi = (\ell \w)^2 \chi \,.
\ee
Here $\Delta$ is given by (\ref{eq:deltaA}).
The retarded Green function at general points $y,y'$ is given by
\be 
G(\omega,y,y') = \frac{f(y)g(y')}{W(\omega)}~, \quad y < y'~,
\ee
and with $f$ and $g$ reversed for $y > y'$. The object $W(\omega) = g \partial_y f - f \partial_y g$ is the Wronskian and is independent of $y$. 
The modes $f$ and $g$ are solutions to the wave equation with $g$ purely ingoing at the cosmological horizon and $f$
`normalizable' near the origin. It is not hard to see, using our above expressions for the modes, that this Green's function is indeed proportional to $B/A$, as we anticipated. In particular, $g \sim A \chi_\text{n.n.} + B\chi_\text{n}$ and $f \sim \chi_\text{n}$. Computing $W(\omega)$ near the origin $x = 0$, we find $W(\omega) \sim A$ and thus near the origin $G(\omega,y,y') \sim (A y + B y^{2l+d-1})/A$. The first term here is independent of $\omega$ and is a contact term. The remaining piece gives the worldline correlator, which is seen to be $\sim B/A$.

The analysis throughout this letter applies also to gravitational perturbations. In general dimensions there are scalar, vector and tensor gravitational modes in de Sitter spacetime. The equations of motion for gauge invariant perturbations can be mapped into a Schr\"odinger form, similarly to our discussion around (\ref{eq:schrodinger}) above for scalar fields. The resulting Schr\"odinger potentials, for scalar, vector and tensor modes respectively, are \cite{Kodama:2003kk, Natario:2004jd}
\bea
V_S & = & \frac{\D (\D-1)}{\sinh^2 y} - \frac{(d-3)(d-5)}{4 \cosh^2 y} \,, \\
V_V & = & \frac{\D (\D-1)}{\sinh^2 y} - \frac{(d-1)(d-3)}{4 \cosh^2 y} \,, \\
V_T & = & \frac{\D (\D-1)}{\sinh^2 y} - \frac{(d+1)(d-1)}{4 \cosh^2 y} \,. 
\eea
These potentials have the same P\"oschl-Teller-like form as for the scalars in (\ref{eq:schrodinger}). There will be an enhanced algebraic structure, corresponding to the linearly realized $SL(2,\mathbb{R})$ of (\ref{eq:linereal}) and (\ref{eq:linereal2}) below, when the second term in the potential is absent. For scalar fields, this requires the conformal coupling $x = \half$. For the gravitons we see that this only occurs for all types of modes if $d=3$, i.e. in four dimensional de Sitter spacetime! There are no tensor modes in four dimensions. In general dimension we have $x = |d-4|/2$ for scalar gravitons, $x=(d-2)/2$ for vector gravitons, and $x=d/2$ for tensor gravitons. Thus for gravitons, $x$ is an integer in odd spacetime dimension $d+1$ and a half integer in even dimensions.

\section{Quasinormal modes and $SL(2,\R)$ algebras}

The Green's function contains information about the physical excitations of the spacetime, the quasinormal modes.
Quasinormal modes satisfy ingoing boundary conditions at the horizon and are normalisable at the origin, typically with complex $\w$. Because these are the same boundary conditions we used to define the solipsistic Green's function, the quasinormal frequencies will give the zeros of the Wronskian and hence the poles of the retarded Green's function (c.f. \cite{Birmingham:2001pj,Son:2002sd}). 
From the Green's function (\ref{eq:aa}), suitably interpreted for integer $l$ as discussed above, the spectrum is immediately found to consist of two families of poles in the lower half complex plane
\be\label{eq:generalpoles}
\ell \w^{\pm}_n = - i \left(2 h_\pm + l + 2n \right) \,, \quad n \in 0,1,2,\ldots \,.
\ee
These and other quasinormal frequencies we will discuss were correctly found by \cite{LopezOrtega:2006my}.
The wavefunctions of these modes are, dropping the spherical harmonic dependence,
\be\label{eq:generalmodes}
\Phi^{\pm}_n = r^{l} f^{- i \ell \w^{\pm}_n/2} {}_2F_1\left(-n,-n \mp x;\frac{d}{2} + l ; \frac{r^2}{\ell^2} \right) e^{- i \w^{\pm}_n t} \,.
\ee
The hypergeometric functions here are Jacobi polynomials.

\subsection{Conformally coupled scalars: $x = 1/2$}

For the case of a conformally coupled scalar, with $x = 1/2$, the two families of quasinormal modes coalesce into a single equally spaced set of modes, consistent with the simplification of the Green's function to a single ratio of gamma functions in (\ref{eq:Ggrav}), where as we noted the poles should be obtained by continuing to non-integer $l$. Thus
\be \label{ccqnm}
\ell {\omega_m} = - i \left(\half (d-1) + l + m \right)~, \qquad m \in 0,1,2,\ldots \qquad (x= \half) \,,
\ee
and the wavefunctions are
\be\label{eq:mmode}
\Phi_m = r^{l} f^{- i \ell \w_m/2} {}_2F_1\left(-\frac{m}{2},\frac{1-m}{2};\frac{d}{2} + l ; \frac{r^2}{\ell^2} \right) e^{- i \w_m t} \,.
\ee
Now that there is a single sequence of evenly spaced modes, it is possible to define raising and lowering operators that move us between all the different modes. Define
\bea
H_+ & = & i \frac{e^{t/\ell}}{\sqrt{f}} \left(r f \frac{\pa}{\pa r} + \ell \frac{\pa}{\pa t} + f \right) \,, \label{eq:linereal} \\
H_- & = & i \frac{e^{-t/\ell}}{\sqrt{f}} \left(- r f \frac{\pa}{\pa r} + \ell \frac{\pa}{\pa t} - f \right) \,. \label{eq:linereal2}
\eea
The final non-derivative term can easily be absorbed in the first term by rescaling the wave functions on which these operators act.
The operators act on the quasinormal modes (\ref{eq:mmode}) as
\be\label{eq:lin}
H_+ \Phi_m = - i m \Phi_{m-1} \,, \qquad H_- \Phi_m = i (m - 2 i \ell \w_m) \Phi_{m+1} \,.
\ee
Furthermore, if we let
\be
H_0 = i \ell \frac{\pa}{\pa t} \,,
\ee
then we recover the conformal algebra
\be
[H_+, H_-] = - 2 i H_0 \,, \qquad [H_\pm,H_0] = \mp i H_\pm \,.
\ee
These operators are related to the standard $SL(2,\mathbb{R})$ generators by
\be\label{eq:map}
D = \half \left(H_- - H_+ \right) \,, \quad K = - H_0 - \half \left(H_+ + H_- \right) \,, \quad H = H_0 - \half \left(H_+ + H_- \right) \,.
\ee
Note in particular for future reference that $2 H_0 = H - K$. These generators satisfy
\be\label{eq:dhk}
[D,H] = - i H \,, \qquad [D,K] = i K \,, \qquad [H, K] = 2 i D \,.
\ee
The $SL(2,\mathbb{R})$ Casimir then turns out to be closely related to the radial part of the wave equation (\ref{eq:wave}).
Acting on solutions to the wave equation we have
\be \label{quadcas}
\left[ -D^2 + \frac{1}{2} \left( H K + K H  \right) \right] \Phi = \left(l + \frac{d-3}{2}\right)\left(l+\frac{d-1}{2}\right) \Phi~.
\ee 
By equating the Casimir with $\D(\D-1)$, we can read off a conformal weight from the above equation
\be\label{eq:special}
\Delta = l + \frac{d-1}{2} \,.
\ee
We can therefore associate each mode $l$ of the scalar field to a `dual' operator with conformal weight $\Delta$.
The weight agrees with the one in (\ref{eq:deltaA}) above that we extracted from the time dependence (\ref{eq:time}) of the retarded Green's function.
Each scalar field will be dual to a tower of operators, similarly to the harmonics of a scalar field on the internal $S^5$ in the established AdS duality for ${\mathcal{N}}=4$ super Yang-Mills theory. 

The quasinormal modes do not directly furnish a unitary representation of the algebra -- they are not normalizable states and do not have real energies. Nonetheless, equation (\ref{quadcas}) shows that the $SL(2,\mathbb{R})$ algebra built from the quasinormal mode raising and lowering operators acts on solutions to the de Sitter wave equation in a unitary lowest weight representation. Previous works using a conformal algebra to construct quasinormal modes include \cite{Chen:2010ik, Chen:2010sn}.

The $SL(2,\mathbb{R})$ symmetry we have uncovered is not an isometry of de Sitter spacetime. However, the generators $H_\pm$ are conformal Killing vectors. Precisely, they are Killing vectors of the conformally rescaled metric
\be\label{eq:ckv}
d\hat s^2 = \frac{1}{r^2} ds^2 = - \frac{f}{r^2} dt^2 + \frac{dr^2}{r^2 \, f} + d\Omega_{d-1}^2 \,.
\ee
This is the metric of AdS$_2 \times S^{d-1}$. The AdS$_2$ is in hyperbolic or `black hole' coordinates.
Thus the $SL(2,\mathbb{R})$ we have found is the remnant of the full symmetry group of a conformally coupled scalar field in de Sitter spacetime that is preserved by our infalling boundary conditions defining the static patch.
We will return to a discussion of the physical interpretation of this symmetry below. Such a hidden conformal symmetry resembles the situation for non-extremal Kerr black holes \cite{Castro:2010fd,Bertini:2011ga}. In that case a close examination of absorption cross-sections in certain `near' regions of the black hole geometry shows that they take the form of a two dimensional conformal field theory. In our case, we do not need to take a `near' region limit to observe the hidden symmetry.

\subsection{General case: $x \neq 1/2$}

Now return to the case of general $x$, with two series of evenly spaced modes (\ref{eq:generalpoles}). We proceed to construct two $SL(2,\mathbb{R})$ algebras that act as ladder operators on each of the two sets of poles independently. Unlike in the special `nested' case of a conformally coupled scalar, the generators will be nonlinear in time derivatives, although still linear in radial derivatives $\pa_r$. The key step is to obtain recurrence relations for the quasinormal modes (\ref{eq:generalmodes}). Thus
\be\label{eq:raise}
H^{\pm}_+ \Phi_n^\pm = - i n \Phi_{n-1}^\pm \,, \qquad H^{\pm}_- \Phi_n^\pm = i (n - i \ell \w_n^\pm) \Phi_{n+1}^\pm \,.
\ee
Rather than working with nonlinear derivatives in time, we will allow the raising and lowing operators to depend explicitly on the mode number $n$ of the wavefunction on which the operators act. We can reintroduce time derivatives if we wish using $\pa_t = - i \w_n^\pm$.
The first order in $\pa_r$ raising and lowering operators achieving (\ref{eq:raise}) are
\bea
H^{\pm}_+ = i e^{2t/\ell} \left(-\frac{1 - i \ell \w^\pm_n}{2 (n \pm x)} r \frac{\pa}{\pa r} - \frac{i \ell \w_n^\pm (i \ell \w^\pm_n-1)}{2 (n \pm x) f} + \frac{i \ell \w_n^\pm (i \ell \w_n^\pm + d-2) + x^2 + 2 l - \left(l + \half d \right)^2}{4(n \pm x)}  \right) \,, \label{nl1}\\
H^{\pm}_- = i e^{-2t/\ell} \left(-\frac{1 + i \ell \w^\pm_n}{d + 2n + 2 l} r \frac{\pa}{\pa r} - \frac{i \ell \w_n^\pm (i \ell \w^\pm_n+1)}{(d + 2n + 2 l) f} + \frac{i \ell \w_n^\pm (i \ell \w_n^\pm - d + 2) + x^2 + 2 l - \left(l + \half d \right)^2}{2 (d + 2n + 2 l)}  \right) \,.\label{nl2}
\eea
Recall once again that $n$ refers to the level of the state upon which these operators act.
We could have also built higher order in $\pa_r$ raising and lowering operators; second order generators will arise
from a systematic construction in the following section. These operators depend explicitly on the angular momentum $l$. Computing the commutator gives a pair of conformal algebras
\be\label{eq:alg}
[H_+^\pm, H_-^\pm] = - i \ell \w_n^\pm = 2 \ell \frac{\pa}{\pa (2 t)} \equiv - 2 i \widetilde H_0 \,, \qquad [H^\pm_\pm, \widetilde H_0] = \mp i H^\pm_\pm \,.
\ee
The Casimir is now
\be \label{quadcas2}
\left[ -D^2 + \frac{1}{2} \left( H K + K H  \right) \right]^\pm  = \left(h_\pm + \frac{l}{2} - 1 \right) \left(h_\pm + \frac{l}{2} \right) ~.
\ee 
Consequently we can read off conformal weights associated to the quasinormal modes under these $SL(2,\mathbb{R})$s as
\be\label{eq:hpm}
\Delta^\pm = h_\pm + \frac{l}{2} \,. 
\ee
Recall that the weights $h_\pm$ were presciently introduced in (\ref{eq:weights}) above.

The formula (\ref{eq:hpm}) associating conformal weights to a scalar field as a function of its mass is reminiscent of analogous formulae appearing in approaches to de Sitter holography based on the future boundary, e.g. \cite{Strominger:2001pn}. Our formula has an extra factor of $\half$ as well as the $l/2$ term due to the fact that different modes on the sphere correspond to different operators in the putative dual $0+1$ dimensional quantum mechanics. Our weights have been read off from an $SL(2,\mathbb{R})$ acting in the static patch rather than at future infinity. Nonetheless, we can observe that in fact the weights of the quasinormal mode generating algebra are directly related to the weights of the fields at future and past infinity.
In \cite{Bousso:2001mw, Anninos:2011jp}, formulae are derived for matching solutions to the de Sitter wave equation between the future, past, `Northern' and `Southern' diamonds. In these formulae, we see that one family of quasinormal modes in the static patch match onto fast decaying modes at future infinity, while the other family of quasinormal modes match onto fast decaying modes as one moves backwards in time towards past infinity. In both cases there is no flux emanating from the antipodal static patch. Thus, the excitations on which the quasinormal mode algebra acts match onto the normalizable states on which the algebras at future and past infinity act.

Unlike the previous case of the conformally coupled scalar, the generators (\ref{nl1}) and (\ref{nl2}) and consequent algebras (\ref{eq:alg})
are not defined on general wavefunctions, but only on each of the two sets of quasinormal modes separately.
Recall that the quasinormal modes are not normalizable states and do not necessarily provide a complete basis for wavefunctions.
These algebras therefore are insufficient to determine the Green's function in the general case.
In the following section we will discuss the general case more systematically, and show how the Green's function descends from the action
of a larger $SL(2,\R) \times SL(2,\R)$ algebra.

\section{$SL(2,\R) \times SL(2,\R)$ and the Green's function}

The P\"oschl-Teller-like potentials arising in our study are in fact known to have a rich algebraic structure seemingly distinct to that we have developed so far. There is an $SO(2,2)$ `potential group' that can be used to obtain the spectrum \cite{Barut:1987am, wuA,wuB}. In this section we show how a version of the formalism discussed in those papers leads to a more satisfactory understanding of the case of general mass, parametrized by $x$. We will introduce an auxiliary AdS$_3$ spacetime together with an $SL(2,\R) \times SL(2,\R)$ algebra that includes a spectrum generating algebra. At the end of this section we explain that the AdS$_3$ arises from a conformal transformation of the static patch of de Sitter spacetime times a line.

Observe that a change of variables maps our P\"oschl-Teller-like Schr\"odinger equation (\ref{eq:schrodinger}) into other Schr\"odinger equations where the potential has the same form but the coefficients are shuffled around. In particular consider the following transformations. Let
\be
z = \log \coth \frac{y}{2} \,, \qquad \chi = \frac{1}{\sqrt{\sinh z}}\psi \,,
\ee
leading to the equation
\be\label{eq:fir}
- \frac{d^2\psi}{dz^2} + \left(\frac{(i \ell \w + \half) (i \ell \w - \half)}{\sinh^2 z} - \frac{(x+ \half) (x- \half)}{\cosh^2 z} \right) \psi =  - (\D - \half)^2 \psi \,.
\ee
Also, we can consider,
\be
z = - 2 i \arctan \tanh \frac{y}{2} \,, \qquad \chi = \frac{1}{\sqrt{i \cosh z}}\psi \,,
\ee
leading to the equation
\be
- \frac{d^2\psi}{dz^2} + \left(\frac{\D (\D-1)}{\sinh^2 z} - \frac{(i \ell \w + \half) (i \ell \w - \half)}{\cosh^2 z} \right) \psi =  - x^2 \psi \,.
\ee
We see that these transformations exchange the places where $\{\w, \D,x\}$ appear in the equation while retaining the P\"oschl-Teller-like form. It will shortly become apparent why this is useful. In the following we will only use the former of the two transformations, and the corresponding equation (\ref{eq:fir}). We included the second transformation for completeness and because it may be the more useful form for other questions closely related to those we will now address.

\subsection{Hyperbolic AdS$_3$ and $SL(2,\R) \times SL(2,\R)$}

Consider the auxiliary AdS$_3$ spacetime
\be
x_1^2 - x_2^2 - x_3^2 + x_4^2 = L^2 \,,
\ee
and parametrize the coordinates by
\be
x_1 \pm x_2 = L \cosh z \, e^{\pm \phi} \,, \qquad x_3 \pm x_4 = L \sinh z \, e^{\pm t/\ell} \,.
\ee
The AdS$_3$ metric in these coordinates is
\be
ds^2 = L^2 \left( - \sinh^2 z \, \frac{dt^2}{\ell^2} + dz^2 + \cosh^2 z \, d\phi^2 \right) \,. 
\ee
This metric is AdS$_3$ in hyperbolic coordinates. The $t$ coordinate will be identified with the de Sitter (static patch) time. The Euclideanized time $t_E = it$ has a natural periodic identification, in analogy to the Euclidean de Sitter time. If we periodically identify the new auxiliary direction $\phi$ then we obtain the non-rotating BTZ black hole. We shall not do this, as $x$ will appear as a (noncompact) hyperbolic label of a lowest weight representation of $SL(2,\R)$. The generators of the geometrical $SO(2,2)$ symmetry of AdS$_3$ can be decomposed into two commuting $SL(2,\R)$ algebras. The generators of these algebras adjusted to our coordinates can be taken to be
\bea
A_\pm & = & i e^{\pm (\phi + t/\ell)} \frac{1}{2}\left( \pm \frac{\pa}{\pa z} - \tanh z \frac{\pa}{\pa \phi} - \coth z \, \ell \frac{\pa}{\pa t} \right) \,, \label{eq:first} \\
A_0  & = & \frac{i}{2} \left(\frac{\pa}{\pa \phi} + \ell \frac{\pa}{\pa t} \right) \,, \\
B_\pm & = & i e^{\mp (\phi - t/\ell)} \frac{1}{2} \left( \pm \frac{\pa}{\pa z} + \tanh z \frac{\pa}{\pa \phi} - \coth z \, \ell \frac{\pa}{\pa t} \right) \,, \\
B_0  & = & \frac{i}{2} \left(- \frac{\pa}{\pa \phi} + \ell \frac{\pa}{\pa t} \right) \,. \label{eq:last}
\eea
These satisfy
\be
[A_\pm,A_0] = \mp i A_\pm \,, \quad [A_+,A_-] = - 2 i A_0 \,, \quad [B_\pm,B_0] = \mp i B_\pm \,, \quad [B_+,B_-] = - 2 i B_0 \,.
\ee
The $SO(2,2)$ Casimir is consequently
\bea
C^{SO(2,2)} & = & - 2 \left( A_0^2 + B_0^2 \right) + A_+ A_- + A_- A_+  + B_+ B_- + B_- B_+ \nonumber \\
& = & 2 \left( C_A^{SL(2,\R)} + C_B^{SL(2,\R)} \right) \,.
\eea
In fact using the representation (\ref{eq:first}) -- (\ref{eq:last}), one further finds that $C_A^{SL(2,\R)} = C_B^{SL(2,\R)}$. That is, we are considering a symmetric representation of $SO(2,2)$. Now let our generators act on wavefunctions
\be\label{wavephi}
\Psi(z, \phi, t) = e^{i p \phi} e^{- i \w t} \frac{\psi(z)}{\sqrt{\sinh 2 z}} \,.
\ee
The Casimir acting on such wavefunctions is then found to satisfy
\be\label{eq:casimir}
- \frac{d^2\psi}{dz^2} + \left(\frac{(i \ell \w + \half) (i \ell \w - \half)}{\sinh^2 z} - \frac{(i p+ \half) (i p- \half)}{\cosh^2 z} \right) \psi =  - \left(1 + 4 C^{SL(2,\R)} \right) \psi \,.
\ee
Comparing this expression with our Schr\"odinger equation (\ref{eq:fir}), and setting the momentum in the $\phi$ direction to be $p = - i x$, we find that the wave equation in de Sitter space descends from an action of $SL(2,\R) \times SL(2,\R)$ with Casimir
\be
C^{SL(2,\R)} = \left(\frac{\Delta}{2} + \frac{1}{4} \right) \left(\frac{\Delta}{2} - \frac{3}{4}  \right) \,,
\ee
for each $SL(2,\R)$. This Casimir associates the de Sitter wave equation with a lowest weight representation of $SL(2,\R) \times SL(2,\R)$ with weights
\be\label{eq:deltatilde}
\widetilde \Delta = \frac{\Delta}{2} + \frac{1}{4} = \frac{l}{2} + \frac{d}{4} \,.
\ee
The wavefunctions (\ref{wavephi}) correspond to the hyperbolic parameterization of lowest weight representations by the continuous spectrum of the $H - K$ generator of the conformal algebra \cite{lindblad}. The structure we have just described fits most directly with the larger mass case of imaginary $x$, so that the momentum $p= - i x$ is real. For low mass scalars and gravitons, with $x$ real, the wavefunctions (\ref{wavephi}) can be thought of as bound states, as we will shortly take a $\Z_2$ gauging of the $\phi$ direction. In all cases we can use the basis (\ref{wavephi}) to parametrize the states and compute the Green's function. We can then evaluate the Green's function at real or imaginary momentum $p$ as appropriate.

The $SL(2,\R) \times SL(2,\R)$ algebra we have just described takes us outside the space of solutions of the de Sitter wave equation for a scalar field of a given mass, because as well as shifting the frequencies $\w$, the generators also move us around in $p = - i x$ space. We must restrict to fixed $x$ to obtain de Sitter physics. We would like to think of this as a type of level matching condition. Furthermore, because the de Sitter equation depends only on $x^2$, we must also introduce a $\Z_2$ gauging of the states that identifies $p \leftrightarrow - p$. The states of interest therefore take the form
\be\label{eq:prodstates}
| \w \rangle = | \half( \ell \w + p) \rangle_A \otimes | \half (\ell \w - p) \rangle_B + | \half(\ell \w - p) \rangle_A \otimes | \half(\ell \w + p) \rangle_B \,.
\ee
While auxiliary from a de Sitter point of view, the existence of two $SL(2,\R)$s acting on the wave equation now fixes the Green's function. This is the nature of `potential algebras', which are symmetries relating different Schr\"odinger equations, and under which the Green's functions must transform. We noted in section \ref{sec:first} above, and will prove in the following section, that the Green's function of a theory with a single $SL(2,\R)$ evolved with respect to the $H-K$ generator falls off like an inverse power of $\sinh t/\ell$ as in (\ref{eq:time}) above. This expression Fourier transforms into a ratio of gamma functions, as we also noted above. It follows that acting on our product states (\ref{eq:prodstates}) with their respective $SL(2,\R)$ actions, the Green's function takes the form
\bea
G^R_\text{dS}(\w) & = & G_{\widetilde \Delta}^R(\half(\ell \w - i x)) G_{\widetilde \Delta}^R(\half(\ell \w + i x)) \nonumber \\
& \propto & \frac{\G\left(\widetilde \Delta - i \half (\ell \w - i x) \right)}{\G\left(1 - \widetilde \Delta - i \half (\ell \w - i x) \right)} \frac{\G\left(\widetilde \Delta - i \half (\ell \w + i x) \right)}{\G\left(1 - \widetilde \Delta - i \half (\ell \w + i x) \right)} \nonumber \\
 & = & \frac{\G\left(\half (l - i \ell \w) + h_+ \right)}{\G\left(\half (2 - d - l - i \ell \w) + h_+ \right)} \frac{\G\left(\half (l - i \ell \w) + h_- \right)}{\G\left(\half (2 - d - l - i \ell \w) + h_- \right)} \,. \label{eq:factored}
\eea
In the final expression we see that we recover precisely the retarded Green's function obtained in (\ref{eq:aa}) above by solving the de Sitter wave equation explicitly.

It is of interest to elucidate more explicitly the nature of the `chiral quarks' that are the constituent states in the tensor product de Sitter state of equation (\ref{eq:prodstates}). These states should have a Green's function given by a single ratio of gamma functions, rather than a product. We shall not explore this question here, but note that a candidate description for such states is as charged particles on finite temperature AdS$_2$ with an electric field. These can have the required Green's functions \cite{Faulkner:2011tm, Spradlin:1999bn}. The idea would be that de Sitter states are neutral composite states of a charged particle and antiparticle. Another context where similar correlators arise, single ratios of gamma functions, is in Liouville theory (e.g. \cite{Nakayama:2004vk} and references therein). Liouville theory is also closely related to two dimensional dilatonic black holes and the $SL(2,\R)/U(1)$ coset model. Many formulae appearing in those discussions are reminiscent of expressions we have encountered here; it would be exciting to uncover a closer connection. The level matching implicit in (\ref{eq:prodstates}) would suggest that we are gauging a hyperbolic generator of $SL(2,\R)$, rather than the parabolic generator that is gauged in the setups we have just described.

The appearance of a left and right moving $SL(2,\R)$ structure, as well as a possible BTZ black hole, naturally makes one wonder about the possible existence of an underlying CFT whose central charge could reproduce the entropy of de Sitter space. We hope to investigate this question in the future. A starting point for this investigation may be the following observation that $dS_{d+1} \times S^1$ is conformal to $\text{BTZ} \times S^{d-1}$:
\bea
ds^2_{dS_{d+1} \times S^1} & = & - \tanh^2z \, dt^2 + \frac{\ell^2}{\cosh^2 z} \left(dz^2 + d \Omega_{d-1}^2 \right)+ \ell^2 d \phi^2 \\
& = & \frac{\ell^2}{\cosh^2 z} \left( - \sinh^2 z \frac{dt^2}{\ell^2} + dz^2 + \cosh^2 z \, d\phi^2 + \Omega_{d-1}^2 \right) \\
& = & \frac{\ell^2}{\cosh^2 z} ds^2_{\text{BTZ} \times S^{d-1}} \,. \label{eq:btz}
\eea
The location of the static observer, $z \to \infty$, maps onto the conformal boundary of $AdS_3$. This may support the notion of imposing boundary conditions on the static patch worldline.

The above conformal map also explains, mathematically at least, how the $SL(2,\R) \times SL(2,\R)$ symmetry is related to the de Sitter wave equation. Starting with a wave equation in de Sitter space, by appending an extra line to the spacetime we can obtain the same equation as that of a conformally coupled field in $dS_{d+1} \times \R$ with a fixed momentum along the line. This equation for a conformally coupled field is now equivalent to that of a conformally coupled field on AdS$_3 \times S^{d-1}$. The $SL(2,\R) \times SL(2,\R)$ algebra then appears as the isometries of the AdS$_3$ factor in this spacetime.

The conformal map (\ref{eq:btz}) has appeared before in different contexts, such as \cite{Hubeny:2009rc}. The strategy of studying de Sitter physics via a conformally related Anti-de Sitter spacetime was advocated in \cite{Alishahiha:2005dj}. In the AdS frame, the conformal factor appears as a spatially dependent Planck mass. That paper also discusses how a 0+1 dimensional description of de Sitter space can emerge by iteratively applying the logic of the dS/dS correspondence \cite{Alishahiha:2004md}. It is of interest to understand better the connection between that quantum mechanical description and the solipsistic approach we are investigating here.

\subsection{Quasinormal modes revisited}

We can furthermore use the above  $SL(2,\R) \times SL(2,\R)$ structure to obtain all the quasinormal modes of the de Sitter wave equation by purely algebraic means. In doing so we also exhibit a different $SL(2,\R)$, similar to that at the end of the previous section, that acts on quasinormal modes. Let us consider the subspace of the algebra that preserves the value of $x$, i.e. such that the $\phi$ dependent exponential factor in the wavefunctions (\ref{wavephi}) remains constant. These are operators that commute with $A_0 - B_0 = i \pa_\phi$, and thereby act on solutions to the wave equation for a given mass of the scalar field. We are led to consider the action of the following generators
\be
J_+ \equiv A_+ B_+ \,, \qquad J_- \equiv A_- B_- \,, \qquad J_0 \equiv A_0+B_0 =  i \ell \frac{\pa}{\pa t} \,.
\ee
The $J_+$ and $J_-$ act as raising and lowering operators for $J_0$ and leave the $\phi$ dependence of the wavefunction invariant.
These operators do not directly furnish an $SL(2,\R)$ algebra. Instead, they satisfy
\be
 [ J_0, J_+] = 2 i J_+ \,, \quad [ J_0 , J_- ] = - 2 i J_- \,,  \quad [J_+, J_-] = \frac{i}{2} J_0^3 + i \left(2 \widetilde \Delta (\widetilde \Delta - 1) + \frac{x^2}{2} \right) J_0 \,.\label{eq:J}
\ee
To obtain the final expression we used the Casimir equation (\ref{eq:casimir}). This implies that the above algebra has been written in a way that is dependent on the representation being considered. $\widetilde \Delta$ is defined in (\ref{eq:deltatilde}).

To generate the quasinormal mode spectrum, following the logic of the previous section, consider wavefunctions satisfying
\be
J_+ \Psi = 0 \,.
\ee
This equation has two types of solution
\be\label{eq:ABsols}
A_+ \Psi^A_0 = 0 \qquad \text{or} \qquad B_+ \Psi^B_0 = 0 \,.
\ee
Each of these equations has a unique solution of the form (\ref{wavephi}), taking e.g. $x = i p > 0$.
We could also take linear combinations of the above equations and in fact a more gauge-invariant description of the following would be to consider solutions to $J_+ \Psi = 0$ that are invariant under $\phi \to - \phi$. The decomposition according to (\ref{eq:ABsols}) will however be more useful. We can therefore define a sequence of wavefunctions by
\be
\Psi^{A,B}_{n} =J_-^n \Psi^{A,B}_{0} \,.
\ee
These wavefunctions have the following spectrum for $J_0$
\be
J_0\Psi^A_n= \ell \omega \Psi_n^A = - i \left(2 \widetilde\Delta + x + 2 n\right)  \Psi_n^A \,,
\ee
\be
J_0 \Psi^B_n= \ell \omega \Psi_n^B = - i \left(2 \widetilde\Delta - x + 2 n\right)  \Psi_n^B \,.
\ee
This is precisely the quasinormal spectrum previously found in (\ref{eq:generalpoles}) by explicitly solving the wave equation. In fact, following through the various changes of variables we have made, we can check explicitly that the $\Psi^{A}_{n}$ and $\Psi^{B}_{n}$ are the same wavefunctions as $\Phi_n^+$ and $\Phi_n^-$ in (\ref{eq:generalmodes}), respectively.

The nonlinear quasinormal mode spectrum generating algebra of the $J$ operators in (\ref{eq:J}) can be turned into an $SL(2,\R)$ algebra by rescaling the generators by terms that are nonlocal in $\pa/\pa t$. In this way, fixing $x$ which is not shifted by this algebra, one obtains generators of a similar form to those we wrote down previously in (\ref{nl1}) and (\ref{nl2}). The difference is that the operators will now be second order in radial derivatives and that we have arrived at the generators via a more systematic algebraic procedure. Define
\be\label{eq:rescale}
H^\pm_+ = \frac{2}{\pa_t \pm x - 2 \widetilde \Delta} \, J_+ \,, \qquad H^\pm_- = \frac{2}{\pa_t \mp x + 2 \widetilde \Delta} \, J_- \,, \qquad \widetilde H_0 = \half J_0 \,.
\ee
By explicit computation or by using (\ref{eq:J}), the above generators are seen to satisfy the $SL(2,\R)$ algebras of equation (\ref{eq:alg}).
The corresponding quadratic Casimirs are again found to be given by (\ref{quadcas2}). The two sets of generators do not commute between themselves, and indeed the $H^+$ algebra is found to generate the $\Phi^+$ quasinormal modes, while the $H^-$ algebra generates the $\Phi^-$ modes, as in (\ref{eq:raise}) above. Thus the two families of quasinormal modes are once again found to be associated with lowest weight representations of $SL(2,\R)$ with weights
\be
\Delta^\pm = \widetilde \Delta \pm \frac{x}{2} = h_\pm + \frac{l}{2} \,.
\ee
As in the discussion below (\ref{eq:hpm}), by matching through the horizon, these weights appear to be related to the weights of states that can be read off at future infinity. This notion of matching through the horizon also suggests an interpretation for the factorized form of the Green's function in (\ref{eq:factored}). The existence of the antipodal static patch suggests that, similarly to other finite temperature spacetimes, the correlation functions may also be obtained from a Keldysh formalism. It is possible that the factorized form (\ref{eq:factored}) can be interpreted as the amplitude to transition from one observer to behind the horizon and then from behind the horizon through to the antipodal observer. We hope to elucidate such interpretational issues in the future. At stake is the intrinsic physical interpretation of the two copies of $SL(2,\R)$ for de Sitter space, beyond the projection to a fixed momentum that is described by the $J$ algebra.

\subsection{Supersymmetric structure}

Extending slightly the algebras we have discussed so far, we can furthermore exhibit a special supersymmetric structure emerging when $x= \frac{1}{2}$. Firstly, introduce in addition the `parity' operator
\be
P: \Psi(z, \phi, t) \mapsto \Psi(z, -\phi, t) \,.
\ee
This $\Z_2$ operator exchanges the $A$ and $B$ generators: $P B P = A$.
Then, it is easily checked that in addition to the $J_\pm$ operators moving us up and down each family of quasinormal modes separately, the operators
\be
Q_- \equiv B_- P = PA_-  \quad \text{and} \quad Q_+ \equiv P B_+ = A_+ P \,,
\ee
lower and raise the quasinormal modes, respectively, while alternating between the two families. Clearly $J_- = Q_-^2$. The structure is illustrated in the following figure \ref{fig:half}.
\begin{figure}[h]
\begin{center}
\includegraphics[width=130pt]{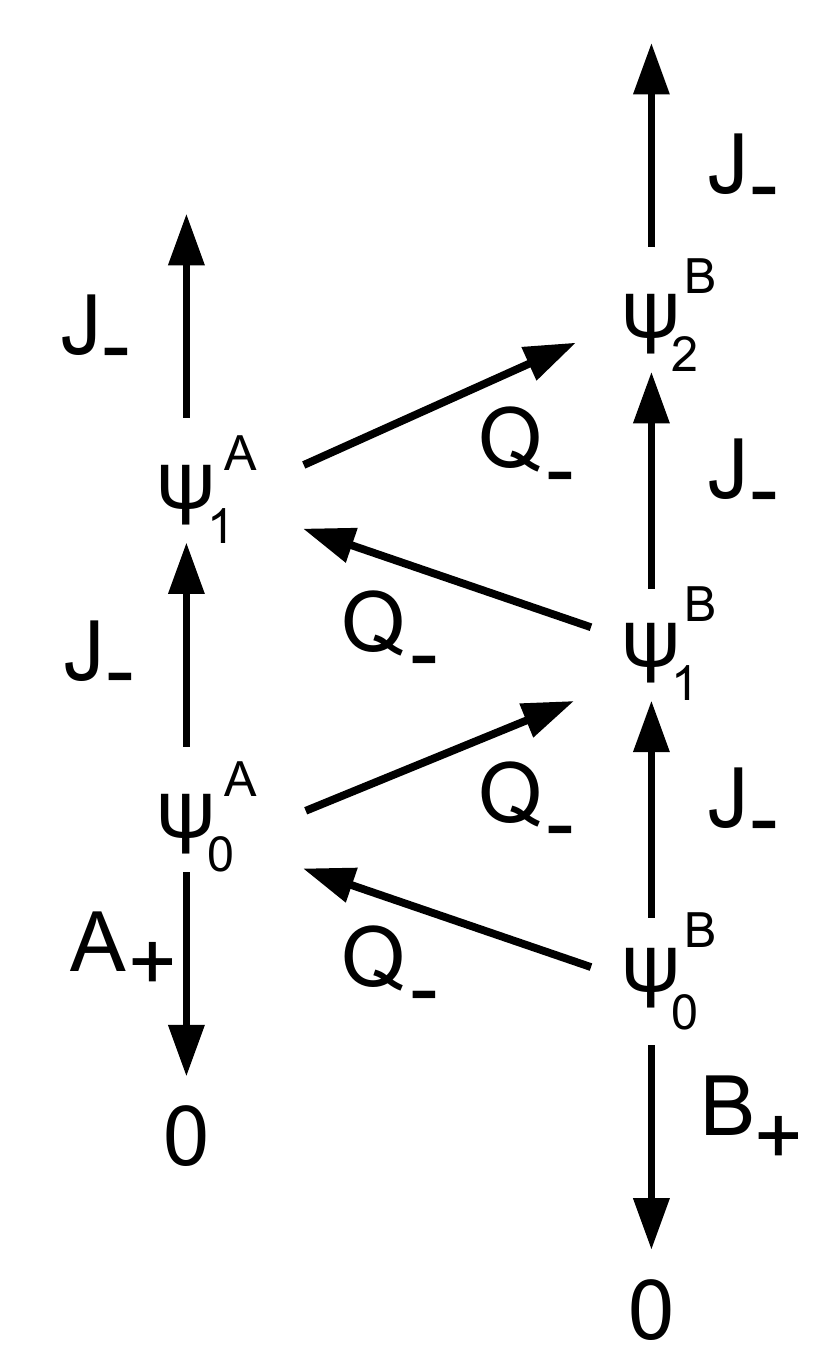}\caption{Quasinormal mode lowering operators when $x = \half$. In addition to the $J_-$ operators, the lowering operator $Q_-$ maps the two families of quasinormal modes into each other. \label{fig:half}}
\end{center}
\end{figure}
While the lowering operator $Q_-$ as defined does not square to zero, one can straightforwardly introduce associated supercharges that do square to zero. This is done by associating a fermion number to one of the families of quasinormal modes and introducing two lowering operators that 
raise or lower the fermion number \cite{Fubini:1984hf}. By combining all the quasinormal modes into one sequence of raising and lowering operators, we have recovered the enhanced algebraic structure observed for $x = \half$ in the previous section.

More generally, the cases of $x$ integer or half integer can be expected to have an enhanced algebraic structure. For instance, for $x$ integer, which  includes the case of gravitons in odd spacetime dimension $d+1$, we see that after allowing for a finite number of low lying states in one of the families, the spectrum of quasinormal modes of the two families becomes degenerate. The degenerate modes nonetheless correspond to different eigenfunctions. This situation is completely analogous to that found in theories of supersymmetric quantum mechanics. In that case, two Hamiltonians, describing the supersymmetric partners, can be obtained from the same superpotential, leading to spectra of the type just described \cite{FernandezC.:2009tc}. In the case of $x$ half integer, the situation is similar in that the algebra can also be enlarged with supercharges. In these cases they do not commute with the Hamiltonian, but nonetheless allow the spectrum to be generated algebraically. An example of a similar system that can be analyzed from this perspective is the simple harmonic oscillator, where an $osp(1|2)$ structure is known to arise \cite{Macfarlane:1991pb}.

The supersymmetric structure of certain P\"oschl-Teller potentials was recently used to show that odd dimensional de Sitter spacetimes are transparent \cite{Lagogiannis:2011st}. That discussion, in global de Sitter space rather than the static patch, did not depend on the mass of the scalar field and so does not map directly onto the structures we are discussing here.

\section{Conformal quantum mechanics}

In the remainder of this letter we describe how the structure of the gravitational correlators we have obtained can be reproduced starting from a dual conformal quantum mechanics theory on the worldline of the static observer. This will lead us to speculate about the nature of possible duals for the de Sitter static patch. The $SL(2,\mathbb{R})$ algebra (\ref{eq:dhk}) can be satisfied by setting
\be
H = i \ell \frac{d}{d\t} \,, \qquad D = i \t \frac{d}{d\t} \,, \qquad K = \frac{i}{\ell} \t^2 \frac{d}{d\t} \,.
\ee
Call $\t$ the scaling time. We are interested in the static de Sitter time. From e.g. (\ref{eq:map})
\be\label{eq:adsds}
i \ell \frac{d}{dt} = H_0 = \frac{1}{2} \left( H - K \right) \,.
\ee
We immediately obtain
\be\label{eq:times}
\t = \ell \tanh \frac{t}{2 \ell} \,.
\ee

The $SL(2,\R)$ invariance constrains the propagators of certain `primary states' in a conformal quantum mechanics theory, which we now define.
Firstly introduce the Cartan operators
\be
L_{\pm} = \pm i D - H_0 \,, \qquad R = \frac{1}{2} \left(H + K \right) \,.
\ee
These satisfy
\be
[R, L_{\pm}] = \pm L_{\pm} \,, \qquad [L_{-}, L_{+}] = 2 R \,.
\ee
Note that these raising and lowering operators are distinct from the $H_\pm$ operators we have considered previously. The $H_\pm$ generators raise and lower the de Sitter energy $H_0$, and act on quasinormal modes, while the $L_\pm$ generators raise and lower $R$, the associated `AdS energy', and act on states. The primary states are eigenstates of the $R$ Hamiltonian that are annihilated by $L_-$
\be\label{eq:conditions}
R | r \rangle = r | r \rangle \,, \qquad L_{-} | r \rangle = 0 \,.
\ee
We now proceed to review how the eigenvalue $r$ determines the propagator of these states. This follows the classic discussion in \cite{de Alfaro:1976je} as well as the more recent paper \cite{Chamon:2011xk}.

Define the new state
\be\label{eq:magicstate}
| r, \t \rangle = N(\t) e^{ - \w(\t) L_{+}} | r \rangle \,,
\ee
parametrised by a label $\t$. Here the time dependent functions are
\be
N(\t) = \sqrt{\G(2 r)} \left(\frac{\w(\t) + 1}{2} \right)^{2 r} \,, \qquad \w(\t) = \frac{\ell + i \t}{\ell - i \t} \,.
\ee
The reason for introducing this state is that a small amount of algebra shows that it satisfies
\bea
H | r, \t \rangle & = & - i \ell \frac{d}{d\t} | r, \t \rangle \,, \\
D | r, \t \rangle & = & - i \left( \t \frac{d}{d\t} + r \right) | r, \t \rangle \,, \\
K | r, \t \rangle & = & - \frac{i}{\ell} \left( \t^2 \frac{d}{d\t} + 2 r \t \right) | r, \t \rangle \,.
\eea
That is, the state describes the time evolution of the state $ |r \rangle$, albeit from the time $\t = i \ell$, and it transforms nicely under $SL(2,\R)$. In particular, some more algebra reveals that it has the scaling correlators
\be\label{eq:niceevolve}
\langle r, \t' | r, \t \rangle = \frac{\G(2 r) \, \ell^{2 r}}{\left(2 i (\t' - \t) \right)^{2 r}} \,.
\ee

To obtain the propagators (\ref{eq:time}), with respect to the de Sitter time (\ref{eq:adsds}) rather than the scaling time, we need to consider a slightly different state, cf. the discussion in \cite{de Alfaro:1976je},
\be
| r, t \rangle_\text{dS} = \left((\t/\ell)^2 - 1\right)^r | r, \t \rangle \,.
\ee
Then, using (\ref{eq:niceevolve}) and the relation (\ref{eq:times}) between the times, 
\be
{}_\text{dS}\langle r, t' | r, t \rangle_\text{dS} = \frac{\G(2 r)}{\left(2 i\right)^{2 r}} \left(\frac{1}{\sinh \half (t-t')/\ell} \right)^{2 r} \,.
\ee
We will see shortly that in the types models we may wish to consider, the ground state $|r_0 \rangle$ of the $R$ Hamiltonian has a large eigenvalue $r_0 \gg 1$. While the proper dual interpretation of this fact remains to be developed, it seems to be consistent with our observation that de Sitter space itself is not invariant under the various $SL(2,\R)$s we discussed, but that they appear as symmetries of the wave equation. The remaining eigenvalues can be written as
\be\label{eq:r0}
r = r_0 + \Delta \,.
\ee
We wish to factor out the time dependence of the background, corresponding very loosely to de Sitter space decaying by `falling through its own horizon', and therefore consider the ratios
\be\label{eq:niceratio}
G^R_\Delta(t) \equiv \theta(t) \frac{{}_\text{dS}\langle r_0 + \Delta, 0 | r_0 + \Delta, t \rangle_\text{dS}}{{}_\text{dS}\langle r_0, 0 | r_0, t \rangle_\text{dS}} =  \theta(t) \left(\frac{r_0}{i \sinh \half t/\ell} \right)^{2 \Delta} \,.
\ee
Here we used the approximation $\frac{\G(2 r_0 + 2 \Delta)}{\G(2 r_0)} \approx (2 r_0)^{2 \Delta}$, valid for $r_0 \gg \Delta$, and we inserted a Heaviside step function in order obtain the retarded propagator. This result (\ref{eq:niceratio}) is the same time dependence that we found for the static patch de Sitter propagator in (\ref{eq:time}) for the case $x = \half$. Dividing the propagator of the excitation by the propagator of the `de Sitter' background itself might be thought as computing the conditional probability for the decay of the mode under the assumption that the background itself continues to exist. 

Consider now the general case with $x \neq \half$. As we noted in passing in section \ref{sec:first}, the general Green's function (\ref{eq:aa}) or (\ref{eq:factored}) is obtained from the Fourier transform of a convolution of Green's functions of the form (\ref{eq:niceratio}). We saw in the previous section that the solipsistic Green's function was controlled by two $SL(2,\R)$ structures. While the constituents are described by a conformal quantum mechanics, in (\ref{eq:prodstates}) we see that the physical de Sitter fields are tensor products of these constituent states with equal de Sitter energies and opposite internal charges $p$. This projection has the flavor of a level matching condition and is similar to the matching imposed in the Kerr/CFT correspondence, see e.g. \cite{Bredberg:2009pv}. In our de Sitter context, it is possible that the need to level match is related to the non-decoupling of gravity on the de Sitter worldline (or possibly an associated worldsheet). This would then be similar to the origin of level matching in string theory from the requirement of invariance under large diffeomorphisms. In terms of these symmetries, we can write the Green's function as
\be\label{eq:convratio}
G^R_{l}(t) = \int dt' \cos{p \, t'} \, G^R_{\widetilde \Delta} (t+ t')  G^R_{\widetilde \Delta} (t- t') \,,
\ee
where we remind the reader that $\widetilde \Delta = \frac{l}{2} + \frac{d}{4}$ and $p = - i x = - i \sqrt{\frac{d^2}{4} - \ell^2 m^2}$. Upon Fourier transforming, this immediately reproduces the result obtained in (\ref{eq:factored}). The momentum $p$ represents the internal energy of the composite state that is being propagated. The case of $x$ real should be interpreted as an analytical continuation of the above result and is associated with a non zero binding energy. The stability bound $x < d/2$ quoted below (\ref{eq:mm22}) then appears as a maximal binding energy. 
As mentioned in the previous section, the appearance of $p$ may be associated with the operation that is necessary to connect the two antipodal static patches via the region behind the horizon.

One challenge that remains is to find a specific conformal quantum mechanics model with the scaling dimensions $\Delta$ given in terms of an $SO(d)$ quantum number $l$ as in e.g. (\ref{eq:special}) or (\ref{eq:deltatilde}). Analogy with established AdS/CFT dualities would suggest that the dual worldline theory should be a large $N$ matrix quantum mechanics. Because the static patch observer experiences a finite temperature, we may expect that the matrix quantum mechanics should be at a finite temperature. The finite temperature free energy of the full matrix theory should reproduce the de Sitter entropy. However, as is also well established, the special property of theories with geometrical gravity duals is that the thermalization of the order $N^2$ off-diagonal matrix elements is manifested in the bulk as an event horizon in a classical spacetime \cite{Witten:1998zw}. It is natural to associate the dynamics of the classical spacetime with the commuting eigenvalues of the matrices (cf e.g. \cite{Banks:1996vh}). In the classical geometric limit, the `stringy' and thermalized off-diagonal excitations have become parametrically heavy compared to the diagonal modes (cf e.g. \cite{Heemskerk:2009pn}). We can then imagine integrating out these modes to obtain an effective action for the eigenvalues. In this effective eigenvalue action, the temperature will appear as a parameter. While the eigenvalues themselves are also at finite temperature, corresponding to the thermal gas of gravitons in de Sitter space, this effect of temperature is subleading in the large $N$ limit compared to the classical dynamics of the eigenvalues governed by their effective temperature-dependent action. 

\subsection{Free theory}

As an (overly simple) example of the types of effective eigenvalue theories we are interested in, consider the free theory of $N$ eigenvalues in $\R^d$
\bea
K = \frac{1}{2 \ell}  \sum_{a=1}^N\vec x^2_a \,, \qquad H =  - \frac{\ell}{2}  \sum_{a=1}^N \nabla^2_a \,, \qquad D = \frac{i}{2} \sum_{a=1}^N  \vec x_a \cdot \nabla_a \,. \label{eq:cqm}
\eea
The primary states in this theory have wavefunctions
\be
\psi(x) = \sum_a c_{i_1 \cdots i_l} x^{i_1}_a \cdots x^{i_l}_a \, e^{- \sum_b \vec x_b^2/(2 \ell)} \,.
\ee
with $c_{i_1 \cdots i_l}$ a symmetric tracefree tensor. The $R$ eigenvalues of these states are $r = dN/2 + l/2$. They therefore lead to conformal dimensions
\be\label{eq:free}
\Delta = \frac{l}{2} \,,
\ee
in the expression (\ref{eq:niceratio}). The dimensions in (\ref{eq:free}) bear comparison with the de Sitter results (\ref{eq:deltatilde}). In particular, the $l$ dependence is exactly correct. The shift of the spectrum by a constant as well as the need to take tensor products of modes remains to be explained.
Alternatively one may try to connect with the special $SL(2,\mathbb{R})$ of conformally coupled scalars and four dimensional gravitons, with scaling dimensions (\ref{eq:special}). Here there is no need to take a tensor product, but one must obtain $\Delta \sim l$ rather than $\Delta \sim l/2$. We hope to investigate the spectrum of interacting conformal eigenvalue dynamics in the future using the collective field method \cite{Jevicki:1979mb}, with a view to reproducing the de Sitter scaling dimensions. It is possible that the need to consider tensor representations that we found for the generic $SL(2,\mathbb{R})$ correlators and spectrum is related to the antipodal tensoring appearing in the construction of de Sitter spacetime in \cite{Parikh:2004wh}.

Many issues remain to be clarified to establish a fully fledged duality between a worldline quantum mechanics and the de Sitter static patch.
Nonetheless, the various hidden conformal structures we have uncovered in this letter give a framework to build upon. Several concrete features in this framework, beyond those we have already stressed, deserve elaboration. Firstly, how evolution under $H_0 \sim H - K$ in conformal quantum mechanics mimics evolution in the presence of horizon by virtue of being unbounded from below (cf. the generators in (\ref{eq:cqm})). Secondly, how the primary states are evolved starting from an imaginary time in (\ref{eq:magicstate}), reminiscent of the Keldysh double contour formalism for finite temperature theories. Thirdly, whether a sharp connection can be made between the fact that the $SL(2,\mathbb{R})$ generators are not isometries of de Sitter spacetime and the fact that the lowest primary state in the conformal quantum mechanics also does not preserve the symmetries of the theory. The connection here may be the way in which the $S^{d-1}$ `emerges' in the quantum mechanics; the conformal rescalings (\ref{eq:ckv}) and (\ref{eq:btz}), that allow the generators to act as isometries, factorize out the $S^{d-1}$. The $SL(2,\R)$ invariant state may be AdS$_2$ or AdS$_3$, with the $S^{d-1}$ emerging as a description of the ground state in the spirit of \cite{Berenstein:2005aa, Berenstein:2008me}.

The notion of a gauge-fixed worldline holography guided by hidden symmetries need not stop with the static patch of de Sitter space. Most closely connected, the static patch Green's functions of maximally rotating de Sitter spacetimes \cite{Anninos:2010gh,Anninos:2009yc,Anninos:2009jt} have a similar pole structure to that of the de Sitter static patch and can presumably be analyzed via the methods we have employed here. Anti-de Sitter space and Minkowski space may also be amenable to an `inside-out' analysis, although in the absence of a cosmological horizon the boundary conditions away from the observer may be less natural. If, following the established language of the AdS/CFT correspondence, we imagine that the spacetime away from the observer corresponds to the low energy limit of the worldline quantum mechanics, then different spacetime asymptopia will be accessed by deformations of the worldline theory by relevant operators. The equivalence principle suggests that the high energy theory describing the immediate vicinity of the observer is universal for all spacetimes, although one must also include the effects of e.g. a finite temperature.

\section{Acknowledgements}

We are pleased to acknowledge helpful conversations with Raphael Bousso, Frederik Denef, Dan Jafferis, John McGreevy, Luiz Santos, Mukund Rangamani, Eva Silverstein and Andy Strominger. S.A.H. and D.M.H. thank the hospitality of the KITP while this paper was finalized. This research was supported in part by the National Science Foundation under Grant No. NSF PHY05-51164, by DOE grant DE-FG02-91ER40654 and the Center for the Fundamental Laws of
Nature at Harvard University. The research of S.A.H. is partially supported by a Sloan research fellowship.

\end{document}